
%
%
\raggedbottom
\magnification=\magstep1
\baselineskip=11pt
\parskip=8pt
\parindent=1.0truecm
\vskip1.0truecm

%
%

%

\def\vs{\vskip 16pt}

\def\pp{\noindent\parshape 2 0truecm 15truecm 2truecm 13truecm}
%
\def\apjref#1;#2;#3;#4 {\par\pp#1, {\it #2}, {\bf #3}, #4. \par}
\def\bpjref#1;#2;#3{\par\pp#1, {\it #2} #3\par}
%
%

%
%


\def\half{{\textstyle{1\over2}}}

\def\Dt{\spose{\raise 1.5ex\hbox{\hskip3pt$\mathchar"201$}}}	
\def\dt{\spose{\raise 1.0ex\hbox{\hskip2pt$\mathchar"201$}}}	
\def\spose#1{\hbox to 0pt{#1\hss}}

\def\lta{\mathrel{\spose{\lower 3pt\hbox{$\mathchar"218$}}
     \raise 2.0pt\hbox{$\mathchar"13C$}}}
\def\gta{\mathrel{\spose{\lower 3pt\hbox{$\mathchar"218$}}
     \raise 2.0pt\hbox{$\mathchar"13E$}}}

%
%
\def\ref#1{$^{#1)}$}

%
%

\parskip=3pt plus1pt minus.5pt

\def\half{{\textstyle {1 \over 2}}}

\def\threehalves{{\textstyle {3 \over 2}}}

\def\Partial#1#2{ {{\partial #1} \over {\partial #2}} }

\def\frac#1#2{{{#1}\over{#2}}}

\def\edth{{\hbox{$\partial$\kern-0.4em\raise0.03ex\hbox{\'{}}
  \kern-0.3em\hbox{}}}}


\def\sqr#1#2{{\vcenter{\vbox{\hrule height.#2pt
      \hbox{\vrule width.#2pt height#1pt \kern#1pt
         \vrule width.#2pt}
      \hrule height.#2pt}}}}

\outer\def\exclaim #1. #2\par{\medbreak
  \noindent{\sl#1.\enspace}{\rm#2}\par
  \ifdim\lastskip<\medskipamount \removelastskip\penalty55\medskip\fi}

\def\({\left(}
\def\){\right)}
\def\<{\left\langle}
\def\>{\right\rangle}

\def\[{\left[}
\def\]{\right]}


\let\text=\hbox

%
%
\centerline{\hfill DAMTP R94/6}
\centerline{\hfill ALBERTA THY/06-94}
\vs
\vs
\centerline{\bf THE ZEL'DOVICH APPROXIMATION AND}
\vs
\centerline{\bf THE RELATIVISTIC HAMILTON-JACOBI EQUATION}
\vs
\vs
\vs
\centerline{\bf D.S. Salopek,$^1$ J.M. Stewart$^2$ and K.M. Croudace$^2$ }
\vs
\vs
\centerline{${}^1$Department of Physics, University of Alberta,
Edmonton, Canada T6G 2J1 }
\vs
\centerline{${}^2$University of Cambridge,
Department of Applied Mathematics and Theoretical Physics}
\centerline{ Silver Street, Cambridge CB3 9EW, England } 
\vs
\vs
\vs
\centerline{ \bf ABSTRACT}
\vs
\noindent
Beginning with a relativistic action principle for the irrotational
flow of collisionless matter, we compute higher order
corrections to  the Zel'dovich approximation by deriving
a nonlinear Hamilton-Jacobi equation for the
velocity potential. It is shown that the velocity of the field may always be
derived from a potential which however may be a multi-valued function
of the space-time coordinates.  In the Newtonian limit,
the results are nonlocal because one must solve
the Newton-Poisson equation.  By considering the Hamilton-Jacobi
equation for general relativity, we set up gauge-invariant
equations which respect causality. A spatial
gradient expansion leads to simple and useful results which are
local --- they require only derivatives of the initial
gravitational potential.

\vs
\vs
\vs
\vs
\centerline{ \it Submitted to }
\centerline{ \it Monthly Notices of the Royal Astronomical Society}
\centerline{ \it February 14, 1994}
\vfill\eject

%
%

\vs
\centerline{ \bf 1. \quad INTRODUCTION}
\vs
Recent redshift surveys of galaxies suggest that  sheet-like structures
are quite common in our Universe. Within a distance of 100$h^{-1}$Mpc,
a Great Wall appears
in the Center for Astrophysics slice of the Universe,
De Lapparent, Geller \& Huchra (1986). A pencil
beam survey, Broadhurst et al (1990) is consistent with
the hypothesis that similar sheets
exist out to a redshift of $z \sim \half$.
The Zel'dovich (1970) approximation  plays a fundamental role
in describing the formation of sheets that arise from the
nonlinear evolution of irrotational collisionless dust
(for a review, consult e.g., Shandarin \& Zel'dovich (1989).
In an attempt to better understand
the formation of cosmic structure,  we show how to
compute higher order terms in the Zel'dovich approximation.
We employ two principal tools: (1) a relativistic action principle
for dust interacting with  gravity, and (2) Hamilton-Jacobi (HJ) theory
used in conjunction with a Taylor series expansion.

Dust with  rest mass $m$ which flows irrotationally
may be described using an
action principle for a dust field $\chi$,
$$
{\cal I}_{dust} = \int d^4x \sqrt{ -g} \bigl  \{ - \half
\rho \( 1+ g^{\mu \nu} \chi_{,\mu} \chi_{,\nu} /m^2 \)  \,
\bigr \}.  \eqno(1.1)
$$
The four-velocity
$$
U^\mu = - g^{\mu \nu } \chi_{,\nu} / m\, . \eqno(1.2)
$$
is proportional to the gradient of the potential $\chi$.
The energy density $\rho$ is a Lagrange multiplier
which ensures that the four-velocity squared is
$-1$:
$$
U^\mu U_\mu = -1 \,  . \eqno(1.3)
$$
Variation with respect to $\chi$ leads to the equation
of conservation
$$
\left ( \rho \chi^{,\mu} \right )_{;\mu} =0 \, ,
$$
whereas variation with respect to the 4-metric $g_{\mu \nu}$
leads to the usual stress-energy tensor:
$$
T_{\mu \nu}= \rho U_\mu U_\nu   \, .
$$
This action principle (1.1) can be used to clarify the analysis
of cosmological systems. It is relativistic and nonlinear.
It will be the focal point of our discussion.

The action for gravity is the standard Einstein-Hilbert form
$$
{\cal I}_{gravity} =  \int d^4x \sqrt{ -g} \; \half\;{}^{(4)}R \,  ,
$$
where ${}^{(4)}R$  is the Ricci scalar of the 4-metric.
Here we have taken $8\pi G=1$.
We shall also assume that $m=1$ which can always be arranged
by rescaling: $\chi \rightarrow m \chi $.
The action principle for dust is a special case of a general
class of fluids considered by Schutz (1970, 1971);
see also Salopek \& Stewart (1992, 1993) who obtain
exact solutions for  a two-fluid system of black-body radiation and
dust.

Hamilton-Jacobi theory facilitates the generation of solutions
for the corresponding field equations.
In section 2, we write down the non-relativistic HJ equation
for a cosmological dust field, Kofman 1991).
The generating function $S$ is essentially a
nonlinear velocity potential which is
closely related to what is directly observable in cosmology (see
e.g., Dekel et al (1990)).
(We assume throughout that, as seen in a Lagrangian description, the
velocity is derivable from a potential.
However when caustics form $S$ may be a multivalued function of the
space-time coordinates.
In an Eulerian description several irrotational flows are superposed
and the mean, Eulerian, flow is no longer irrotational.)
The Zel'dovich approximation is the first two terms
in our Taylor series expansion.
Our technique clarifies the method employed
by Buchert (1989, 1992, 1993) and Moutarde et al (1991).
However, the higher order terms in the series are difficult
to treat analytically because one must solve the Newton-Poisson
equation, which is non-local.

In section 3, we bypass some of the difficulties introduced by
the Newton-Poisson equation by employing the
{\it relativistic Hamilton-Jacobi equation} which
is a functional differential equation. Although the formation of structure
in a cosmological setting is basically non-relativistic,
the relativistic equations are useful because they maintain causality
(which is not true for the Newtonian theory). The generating functional
${\cal S}$ is solved using a spatial gradient expansion.
By choosing our time hypersurface to be a slice where the
dust-field is uniform,  we are able to integrate the
resulting equations using an  iterative technique.
In this way, we solve the energy and the momentum constraints
of general relativity. The energy constraint is essentially
the $G^0_0$ Einstein equation; it is the relativistic generalization
of the Newton-Poisson equation. The $G^0_i$ Einstein equation
yields the momentum constraint which ensures that the
theory is invariant under reparameterizations of the spatial coordinates
(``gauge-invariance'').

Our spatial gradient expansion is similar to solving Einstein's equations using
a Taylor series, an approach that was initially pioneered by
Lifshitz \& Khalatnikov (1964). They were interested in describing
the initial singularity of the big bang
(see also Landau \& Lifshitz (1975)).
However their program was incomplete in that they
did not explicitly solve the  momentum constraint equation, which we solved
recently under quite general conditions (Croudace, Parry, Salopek \&
Stewart (1994) hereafter known as CPSS; see also
Salopek, Stewart, Croudace \& Parry (1994)).
Using the first two terms in our improved gradient expansion,
we generalized the Zel'dovich approximation to
a relativistic framework.
We also investigated higher order terms which involve
only derivatives of the initial gravitational potential ---
the results are local in nature. In this way,
we hope to better understand the formation of sheet-like structures
in the Universe.

Matarrese et al (1993)
considered an alternative relativistic approach to the formation
of cosmic structure. They
simplified and approximated Einstein's equations 
by neglecting the magnetic part of the Weyl tensor.
The resulting equations were local in that they
did not require any information about neighboring points.
They determined an exact solution and they also
integrated the equations numerically.
CPSS demonstrated that the exact solution
of Matarrese et al (1993) corresponded to the exact
Szekeres solution of Einstein's equations.
They showed further that this solution was unstable.
Bertschinger \& Jain (1994) suggested that the
outcome of this instability was cigar
shaped objects which were more generic than sheets.
Counter to one's intuition, they also claimed
that nonlinear structures can form in void regions.
However, although this approach looks promising,
it is still not clear when the
approximation of neglecting the  magnetic part of
the Weyl tensor  breaks down.

In section 4, we compare the results of our gradient expansion
with two exact solutions of Einstein's equations,
namely planar geometries and spherical geometries
(Tolman-Bondi solutions). In general, the higher order
corrections to the Zel'dovich are  significant.
In section 5 we compare our results with the second order theory of
Buchert \& Ehlers (1993), and in the final section
we discuss our results and state our conclusions.

\vs
\centerline{\bf 2. \quad EVOLUTION OF NONLINEAR VELOCITY POTENTIAL}
\vs

The Newtonian limit is essentially a non-relativistic
approximation to general relativity.
In special relativity the energy of a particle with rest mass $m$
is related to its momentum through
$$
E= m \sqrt{ 1 + p^2/ m^2} \, .
$$
The non-relativistic limit
is the first two terms in the Taylor series (in $p/m$) which yields
$$
E= m+ { p^2 \over 2 m } .
$$
In this section, we will use a similar technique to
derive the cosmological HJ
equation for collisionless dust.

\vs
\centerline{\bf 2.1 \quad Non-relativistic Hamilton-Jacobi Equation}
\vs
In a cosmological setting, the dynamics of the matter is
contained in
the scalar evolution equation (1.3) for the dust field:
$$
\big ( \dot \chi- N^i \chi_{,i}\big )/ N =
\sqrt{ 1+  \chi_{|i}  {\chi}^{|i} } \, .
\eqno(2.1a)
$$
Here we have decomposed the metric  using the Arnowitt-Deser-Misner (ADM)
formalism (see, e.g., Misner et al (1973)):
$$
g_{00} \, = \, -N^2 + \gamma^{ij} N_i N_j, \quad
g_{0i} \, = \, g_{i0} = N_i, \quad
g_{ij} \, = \, \gamma_{ij}. \eqno(2.1b)
$$
$N$, $N_i$ and $\gamma_{ij}$ are the lapse function, the
shift function, and the 3-metric, respectively.
A vertical bar $|$ denotes the covariant derivative with respect to
the 3-metric $\gamma_{ij}$.

We now assume that spatial gradients are small and we approximate the
right-hand side of eq.(2.1a) by the
first two terms of a binomial series:
$1 +\half\chi_{|i}  {\chi}^{|i} $.  
We also work with a longitudinal gauge metric:
$$
N= 1- \Phi_H(t, x) , \quad N^i=0, \quad
\gamma_{ij}= a^2(t) [ 1+ 2 \Phi_H(t, x) ] \delta_{ij} \, ;
\eqno(2.2)
$$
here $a(t)= t^{2/3}$ is the scale factor, and
$\Phi_H(t, x)$  is the gauge-invariant variable of
Bardeen (1980) (see also Stewart (1990)).  The $x^i$ are
comoving spatial coordinates.
In cosmological problems which are described using longitudinal gauge,
the deviations of the
metric from Friedman-Robertson-Walker are of the order
of $10^{-5}$, and a linear treatment of the metric
is justified (this is not true
in alternative gauges such as synchronous gauge; see
section 3).  Next, we define the generating function $S$ through
$$
\chi(t,x) = t - S(t,x)
\eqno(2.3)
$$
which leads to  the {\it non-relativistic Hamilton-Jacobi equation} for
a particle moving in a time-dependent potential $- \Phi_H(t,x)$:
$$
0= { \partial S \over \partial t}
+ { 1 \over 2 a^2} {\partial S \over \partial x^i}
{\partial S \over \partial x^i} - \Phi_H\ .
\eqno(2.4a)
$$
All the terms that were dropped are negligible.
The nonrelativistic HJ equation was derived earlier by Kofman (1991)
who noted that it was similar to the eikonal equation
in geometric optics.
If for the sake of illustration the  potential $\Phi_H$ is fixed,
then this equation will exhibit caustics ---
i.e., $S$ will be multivalued. However,  this does not prevent
one from solving the equation (Goldstein (1981)).

The definition of the velocity potential and the four-velocity,
$$
U^\mu \equiv {dx^\mu \over  d \tau} \equiv - g^{\mu \nu} \chi_{,\nu},
$$
(where $\tau$ is proper time)
implies that the coordinate velocity
of a particle is related to $S$ through
$$
{dx^i \over d t}= { 1 \over a^2} { \partial S \over \partial x^i} \, .
\eqno(2.4b)
$$
Hence, the generating function $S$ is essentially a velocity
potential.  Finally, the standard analysis of small perturbations of
the metric leads to the Newton-Poisson relation
$$
-{ 2 \over a^2}   \nabla^2 \Phi_H =
\left( \rho- { 4 \over 3 \chi^2} \right ) \, . \eqno(2.4c)
$$
The right-hand side is just the
density perturbation $\delta \rho_{co}$ on a comoving slice.
In general, $S$ and
hence $\rho$ will be multivalued functions, in which case the
right-hand is summed over all particles with the same spatial
coordinate $x^i$. Finally, far within the Hubble radius, it
is sufficient to approximate $\chi \sim t$ in the last equation.

It is easy to see that these equations are equivalent to
the standard Newtonian equations. If we differentiate eq.(2.4a)
with respect to $x^j$, one finds that
$$
0= { \partial \over \partial t}
\left( { \partial S \over \partial x^j} \right )
+ {d x^i \over d t} { \partial^2 S \over \partial x^i \partial x^j}
-  { \partial \Phi_H \over \partial x^j}, \eqno(2.5)
$$
which may be rewritten using a convective derivative,
$$
{d \over  dt} \equiv { \partial \over \partial t} + { d x^i \over dt}
{ \partial \over \partial x^i}, \eqno(2.6)
$$
leading to
$$
0= { d  \over d t} \left ( { \partial S \over \partial x^j} \right )
-  { \partial \Phi_H \over \partial x^j} \, . \eqno(2.7)
$$
The physical displacement of a particle is denoted by
$$
r^i = a x^i \, ,
$$
and its velocity is
$$
{ d  \over d t} ( a x^i ) =  { \dot a \over a} r^i +
a {dx^i \over d t}\, .
$$
In order to compute the peculiar velocity $v^i$, one
subtracts the velocity,  $(\dot a / a) r^i$, of the Hubble flow,
$$
v^i = { d  \over d t} ( a x^i ) -  { \dot a \over a} r^i
= { 1 \over a } { \partial S \over \partial x^i} \, .
$$
Substituting in eq.(2.7), one obtains the standard cosmological
equation (Peebles (1980)):
$$
0= { d v^i \over d t} + { \dot a \over a} v^i
-  {1 \over a } { \partial \Phi_H  \over \partial x^i} \, . \eqno(2.8)
$$
The density is computed from the equation of continuity
which implies that
$$
\rho(t,x)= \rho_0(t)\, /
\text{det}\( { \partial x^i \over \partial q^j}  \) ,
\eqno(2.9)
$$
where  $q^j$ denotes the initial (Lagrangian) coordinate,
and $\rho_0(t)= 4/ (3t^2)$ is the energy density at
early times.

It is reasonable to assume that one does not commit any gross error
in assuming that the velocity is derived from a potential
$S$. However, one must be aware that the velocity potential may become
multivalued, and even discontinuous.
It is interesting to compare our equations
with the standard linear perturbation analysis of the velocity
potential where one neglects the second term in eq.(2.4a) which is
a quadratic perturbation. This truncated system
is a poor approximation for the formation of cosmological
objects because the virial theorem implies
that for a bound system the kinetic energy is of the order
of the potential energy.

Equations (2.4) are cumbersome to solve directly since
we must solve $S$ as a function of $x^i$ which in turn is
a function of the Lagrangian variables $q^j$.
We will now rewrite them in a form that is more
in character with the HJ formalism that we are developing.
We begin by inverting the spatial  variables: we assume that
the Lagrangian coordinates $q^j$ are functions
of the Euler coordinates $x^i$. The evolution equations for the
Lagrangian coordinates are easy to write  since they
are constants in time: $  dq^j / dt =0$.
As a result, one can write eqs.(2.4) in the form
$$
\eqalignno{
0&= { \partial S \over \partial t}
+ { 1 \over 2 a^2} {\partial S \over \partial x^i}
{\partial S \over \partial x^i} - \Phi_H \, ,
&(2.10a) \cr
0&= {\partial q^j \over \partial  t}+
{ 1 \over a^2} { \partial S \over \partial x^i}
{ \partial q^j \over \partial x^i}\, ,
&(2.10b)  \cr
-{ 2 \over a^2}   \nabla^2 \Phi_H& =
{ 4 \over 3 t^2}
\left( \text{det}\( { \partial q^j \over \partial x^i} \) - 1\right )
\, . &(2.10c) \cr }
$$
In this way all the dependent variables are functions
of the Eulerian coordinates $x^i$.

\vs
\centerline{\bf 2.2 \quad Series Solution of Non-relativistic HJ Equation}
\vs
We will now solve these cosmological equations by making
a Taylor series expansion
in the longitudinal time variable $t$:
$$
\eqalignno{
S(t,x)&= t \, S^{(0)}(x)  + t^{5/3} \, S^{(1)}(x) + t^{7/3}\,  S^{(2)}(x)+
\ldots\, , &(2.11a) \cr
q^j(t,x)&= x^j + t^{2/3} \, q^{(0)j}(x) + t^{4/3} \, q^{(1)j}(x)+ \ldots \, ,
&(2.11b) \cr
\Phi_H(t,x)&= \Phi_H^{(0)}(x) + t^{2/3} \, \Phi_H^{(1)}(x) +
t^{4/3} \, \Phi_H^{(2)}(x)+ \ldots \, . &(2.11c) \cr }
$$

Collecting common terms we find
$$
\eqalignno{
0&= S^{(0)}- \Phi_H^{(0)} \, , &(2.12a) \cr
0&= { 5 \over 3 } S^{(1)} +
{ 1 \over 2 } { \partial S^{(0)} \over \partial x^i}
{ \partial S^{(0)} \over \partial x^i}
- \Phi_H^{(1)} \, , &(2.12b) \cr
0&= { 7 \over 3 } S^{(2)} +
{ \partial S^{(0)} \over \partial x^i}
{ \partial S^{(1)} \over \partial x^i}
- \Phi_H^{(2)} \, , &(2.12c) \cr }
$$
$$
\eqalignno{
0&= { 2 \over 3} q^{(0)j}+ { \partial S^{(0)} \over \partial x^j} \, ,
&(2.13a) \cr
0&= { 4 \over 3} q^{(1)j}+ { \partial S^{(1)}\over \partial x^j}
+ { \partial S^{(0)}\over \partial x^i}
{\partial q^{(0)j} \over \partial x^i} \, , &(2.13b) \cr
0&= 2  q^{(2)j}+ { \partial S^{(2)}\over \partial x^j}
+ { \partial S^{(1)}\over \partial x^i}
{\partial q^{(0)j} \over \partial x^i}
+ { \partial S^{(0)}\over \partial x^i}
{\partial q^{(1)j} \over \partial x^i} \, , &(2.13c) \cr }
$$
$$
\eqalignno{
\nabla^2 \Phi_H^{(0)}  &= -{2 \over 3 }
{\partial q^{(0)i} \over \partial x^i}  \, , &(2.14a)\cr
\nabla^2 \Phi_H^{(1)}  &= -{2 \over 3 }
{\partial q^{(1)i} \over \partial x^i}
-{ 1 \over 3 }
{\partial q^{(0)i} \over \partial x^i} {\partial q^{(0)j} \over \partial x^j
}
+{ 1 \over 3 }
{\partial q^{(0)i} \over \partial x^j} {\partial q^{(0)j} \over \partial x^i}
\, . &(2.14b) \cr }
$$

The solution for the zeroth order terms is
$$
S^{(0)}= \Phi_H^{(0)}= f(x) \, , \quad
q^{(0)j}= -{ 3 \over 2 } { \partial f \over \partial x^j } \, . \eqno(2.15a,b)
$$
The arbitrary function of  the spatial coordinates, $-f(x)$,
may be interpreted as the initial gravitational potential.
The first order terms are expressed in terms of the {\it non-local}
quantity $W$
$$
W= \nabla^{-2} \left[ {\partial^2 f \over \partial x^k \partial  x^l}
{\partial^2 f \over \partial x^k \partial  x^l}
-  \left (\nabla^2 f \right )^2 \right ], \eqno(2.16a)
$$
as
$$
\eqalignno{
S^{(1)}(x)&= -{ 3 \over 4 } { \partial f \over \partial x^i}
{ \partial f \over \partial x^i}  + { 9 \over 14} W \, , &(2.16b) \cr
\Phi^{(1)}_H(x) &= -{ 3 \over 4 } { \partial f \over \partial x^i}
{ \partial f \over \partial x^i}  + { 15 \over 14} W \, , &(2.16c)\cr
q^{(1)j}(x)&= { \partial \over \partial x^j}
\left [ { 9 \over 8 } { \partial f \over \partial x^i}
{ \partial f \over \partial x^i}  - { 27 \over 56} W
\right ] \, . &(2.16d) \cr }
$$

The Zel'dovich approximation is obtained by solving for
$x^i$ in favour of the Lagrangian coordinate $q^k$
using an iterative technique in eq.(2.11b):
$$
x^i = q^i+ { 3 \over 2 } t^{2/3} { \partial f \over \partial q^i}
+ { 27 \over 56 } t^{4/3} \, { \partial \over \partial q^i}
\nabla^{-2} \left[ {\partial^2 f \over \partial q^k \partial  q^l}
{\partial^2 f \over \partial q^k \partial  q^l}
-  \left (\nabla^2 f \right )^2 \right ] \, . \eqno(2.17)
$$
Here $f$ and its derivatives are evaluated at the
point $q^j$ (which is independent of time t) rather than at
$x^i$ which was the case for eqs.(2.16).
The first two terms are the usual Zel'dovich approximation,
and the third term is essentially the improvement suggested
by Buchert and Ehlers (1993).
Here HJ methods simplify the derivation.
For a planar gravitational system, $f(q^i) \equiv f(q^3)$,
where the initial gravitational potential is a function
only of the third spatial coordinate, the matrix
of second derivatives for $f$ is diagonal with a single non-zero component
$$
{ \partial ^2 f \over  \partial q^i \partial q^j}=
{\rm diag} \left [ 0,0, { \partial ^2 f \over  \partial {q^3}\partial
{q^3}}
\right], \eqno(2.18)
$$
and the last term in eq.(2.17) vanishes; hence the Zel'dovich
approximation is essentially exact.
The density is found using eq.(2.9). For a planar
geometry, it is
$$
\rho = { 4 \over 3 t^2 }  { 1 \over
\left | 1 + { 3 \over 2 } t^{2/3}
\left( { \partial ^2 f /  \partial {q^3}\partial {q^3} }  \right)
\right | }\, . 
\eqno(2.19)
$$

In general, further analytic progress is hampered by the non-local
operator $\nabla^{-2}$ appearing in eq.(2.17).

\vs
\centerline{\bf 3. \quad RELATIVISTIC HAMILTON-JACOBI THEORY}
\vs
Using a general relativistic formulation, we show how to
compute the
higher order terms in the Zel'dovich approximation
which describes cosmological collapse.
We utilize the Hamilton-Jacobi equation for
general relativity which is a functional differential
equation that was first written by Peres (1962).
We evolve the 3-metric in a spatial gradient expansion
which is gauge-invariant.
Our method is a significant advance over a Newtonian theory because it is
local at each order. Using the improved Zel'dovich
approximation, we compute the epoch of collapse.

Although the HJ formalism for general relativity
can be applied to arbitrary
time-hypersurfaces, in this section we will restrict
ourselves to comoving surfaces where the velocity
potential $\chi$ is uniform. The line element is then
$$
ds^2 = -d \chi^2 + \gamma_{ij}(\chi, q) dq^i dq^j   \, .
$$
This choice is referred to as comoving, synchronous
gauge, in which case the 3-metric $\gamma_{ij}$ has physical
significance: it gives the physical distance,
$ds = (\gamma_{ij} dq^i dq^j)^{1/2}$,
between two comoving observers separated by an infinitesimal
comoving distance $dq^i$.  $\chi$ is related to the longitudinal
time variable $t$ through eq.(2.3).

The {\it relativistic Hamilton-Jacobi equation}
for the restricted generating functional
${\cal S}[\gamma_{ij}(x)| \chi ]$ is
$$
0=
{ \partial { \cal S}  \over \partial  \chi } +
\int d^3x \left \{  \, \gamma^{-1/2}
{ \delta { \cal S} \over \delta \gamma_{ij}(x) }
{ \delta { \cal S} \over \delta \gamma_{kl}(x) } \,
 \bigl[  2 \gamma_{jk}(x) \gamma_{il}(x)
- \gamma_{ij}(x) \gamma_{kl}(x) \bigr ]
-{ 1 \over 2 }\gamma^{1/2} \, R  \right \} .  \eqno(3.1)
$$
It is a functional differential equation.
In addition one must satisfy the
the momentum constraint
$$
0 = -2 \left ( \gamma_{ik}
{ \delta { \cal S} \over \delta \gamma_{kj}(x) } \right )_{,j}
+ { \delta { \cal S} \over \delta \gamma_{lk}(x) } \gamma_{lk,i} ,
\eqno(3.2)
$$
which states that the restricted generating functional
${\cal S}[\gamma_{ij}(x)| \chi ]$
remains invariant under reparametrizations of the spatial
coordinates. Eq.(3.1) is essentially the $G^0_0$
Einstein equation --- it is the relativistic generalization
of the Newton-Poisson equation (2.4c). Eq.(3.2) is the $G^0_i$
Einstein equation, which enforces gauge invariance.
If eqs.(3.1) and (3.2) have been solved
for the generating functional ${\cal S}$,
then all of Einstein's equations are satisfied provided
that the 3-metric evolves according to
$$
{ \partial \gamma_{ij} \over \partial \chi }
=  2  \gamma^{-1/2}\,
{ \delta { \cal S}\over \delta \gamma_{kl}(x)}\,
  \biggl ( 2  \gamma_{ik} \gamma_{jl} - \gamma_{ij} \gamma_{kl}
  \biggr ) \, . \eqno(3.3)
$$
This is a first-order nonlinear differential equation for
$\gamma_{ij}$.

\vs
\centerline{\bf 3.1 Gradient Expansion for Generating Functional }
\vs
We look for a solution of the relativistic HJ equation (3.1) in the form
of a spatial gradient expansion
$$
{\cal S}= {\cal S}^{(0)} + {\cal S}^{(2)} + {\cal S}^{(4)}+ ... \, ,
\eqno(3.4)
$$
where
$$
\eqalignno{
{\cal S}^{(0)}=&  - 2 \int d^3x \gamma^{1/2} H(\chi) \, , &(3.5a)\cr
{\cal S}^{(2)}=&  \int d^3x \gamma^{1/2} J(\chi) R \, , &(3.5b)\cr
{\cal S}^{(4)}=&  \int d^3x \gamma^{1/2} \left [ L(\chi) R^2 + M(\chi)
R^{ij} R_{ij} \right ] \, . &(3.5c) \cr }
$$
${\cal S}^{(0)}$ contains no spatial gradients whereas
${\cal S}^{(2)}$ contains two spatial gradients, and so on.
These functionals are invariant under transformations of the
spatial coordinates since they are constructed using the
invariant volume $ d^3x \, \gamma^{1/2}$ and the
Ricci tensor $R_{kl}$ of the 3-metric $\gamma_{ij}$.
$H, J, L, M$ are arbitrary functions of the velocity potential
$\chi$ which are chosen so as to satisfy the relativistic HJ equation
order by order:
$$
H^2 + { 2 \over 3 } { d H \over d \chi }=0 \, ,
\qquad {\rm (zeroth \; order)} \eqno(3.6a)
$$
$$
{ d J \over d \chi} + JH- { 1 \over 2 } = 0 \, ,
\qquad {\rm (second \; order)} \eqno(3.6b)
$$
$$
\eqalignno{
& { d L \over d \chi} -LH- {3 \over 4 } J^2 =0 \,
, \qquad {\rm (fourth \; order)} &(3.6c) \cr
& { d M \over d \chi} -MH+ 2 J^2 =0 \, .
\qquad {\rm (fourth \; order)} &(3.6d) \cr }
$$
As a result, the functional differential equation (3.1) has been
reduced to a series of ordinary differential equations (3.6)
which possess the trivial solution:
$$
H= {2 \over 3 \chi} \, , \quad {\rm (zeroth \; order)},\eqno(3.7a)
$$
$$
J= { 3 \over 10} \chi \, , \quad {\rm (second \; order)}, \eqno(3.7b)
$$
$$
L= { 81 \over 2800} \chi^3 \, , \quad M= -{ 27 \over 350} \chi^3 \, ,
\quad {\rm (fourth \; order)}.
\eqno(3.7c,d)
$$
The constant of integration in (3.7a) can be suppressed by choosing
the origin of $\chi$ appropriately.
The constants of integration for (3.7b--d) become irrelevant at large
$\chi$ and have been ignored.
Using HJ theory, we have thus been able to solve the
constraints that appear within Einstein's equations.

\vs
\centerline{\bf 3.2 Gradient Expansion for 3-metric}
\vs
We solve the evolution equation for the
3-metric by using an iterative technique.
If we neglect all second order spatial gradients
and replace ${\cal S}$ by the zeroth order
approximation ${\cal S}^{(0)}$ in eq.(3.3)
we find that the 3-metric evolves according to
$$
\gamma^{(1)}_{ij}(\chi, q) = \chi^{4/3} \, k_{ij}(q) \, , \ \
\eqno(3.8)
$$
which is accurate to first order in spatial gradients.
This is the long-wavelength approximation.
The `seed metric' $k_{ij}(q)$ is an arbitrary function
that is independent of time;
it describes the initial fluctuations whose wavelengths
are larger than the Hubble radius. This result is very
old --- it was known to Lifshitz \& Khalatnikov (1964).

We thus derive the higher order corrections in a systematic way.
For example, accurate to third
order in spatial gradients, we have found that
$$
\gamma^{(3)}_{ij}(\chi, q) =
\chi^{4/3} \left [  k_{ij}(q) +
{ 9 \over 20 } \chi^{2/3}  \,
\left (  \hat R(q) \,  k_{ij}(q) - 4 \hat R_{ij}(q)   \right )
\right ],  \ \, \quad {\rm (3rd \ order)} \eqno(3.9)
$$
and to fifth order, we obtained
$$
\eqalignno{
&\gamma^{(5)}_{ij}(\chi, q)  =
\chi^{4/3} k_{ij} + { 9 \over 20 } \chi^2 \,
\left [  \hat R \,  k_{ij} - 4 \hat R_{ij}   \right ]+
\qquad  \qquad \qquad \qquad  {\rm (5th \ order)} &(3.10) \cr
& { 81 \over 350 } \chi^{8/3} \Bigg [
k_{ij}\left ( - 4 \hat R^{lm} \hat R_{lm}+
{89 \over 32}  \hat R^2 + { 5 \over 8 } \hat R_{;k}{}^{;k} \right )
 + { 5 \over 8 } \hat R_{;ij} - 10 \hat R \hat R_{ij} +
17 \hat R^l{}_i \hat R_{lj}
- { 5 \over 2 } \hat R_{ij;k}{}^k \Bigg ].   }
$$
Here $\hat R_{ij}(q)$ is the Ricci tensor of the seed metric,
and a semi-colon (;) denotes its corresponding covariant derivative.
(For further details, see CPSS.)
Using an alternative method, these expressions have been
verified by Comer et al (1994).

\vs
\centerline{\bf 3.3 Improved Gradient Expansion for 3-metric}
\vs
Unfortunately, as time increases,
the third order expression, eq.(3.9),
leads to nonsensical results: the determinant of the
3-metric can actually become
negative.  A similar problem occurs when one naively
approximates a non-negative function ${\rm cos}^2(x) \sim 1- x^2$
with the first two terms of a
Taylor expansion --- the approximate function falls below
zero when $x > 1$. One can easily remedy this problem by expanding
${\rm cos}(x) \sim 1- x^2/2$ , and then approximating ${\rm cos}^2(x)$
by the square of this result --- this technique guarantees
a positive result. In an analogous way,
we can improve the expansions (3.9) and (3.10) by expressing the results
as a `square.' The improved 3-third order result
$$
\tilde \gamma^{(3)}_{ij}(\chi,q) =
\chi^{4/3} \
\left \{ k_{il} + { 9 \over 40 } \chi^{2/3}
\left [ \hat R k_{il} - 4 \hat R_{il} \right ] \right \} \,
k^{lm} \,
\left \{ k_{jm} + { 9 \over 40 } \chi^{2/3}
\left [ \hat R k_{jm} - 4 \hat R_{jm} \right ] \right \},  \eqno(3.11)
$$
yields the relativistic generalization of the
Zel'dovich approximation (see section 3.4).
The improved fifth order result is
$$
\eqalignno{
\tilde \gamma^{(5)}_{ij}(\chi,q) & =
\chi^{4/3}
 \left \{ k_{il} + { 9 \over 40 } \chi^{2/3}
\left [ \hat R k_{il} - 4 \hat R_{il} \right ] +
\chi^{4/3} F_{il} \right \} \,
k^{lm} \,  \cr
 &\left \{ k_{jm} + { 9 \over 40 } \chi^{2/3}
\left [ \hat R k_{jm} - 4 \hat R_{jm} \right ] +
\chi^{4/3} F_{jm}\right \}  \, ,  &(3.12a) }
$$
where
$$
F_{ij}  =  { 81 \over 5600}
\Bigg \{
 k_{ij} \left [ -32 \hat R^{lm} \hat R_{lm} + 5 \hat R_{;k}{}^{;k} +
{ 41 \over 2 }  \hat R^2 \right ]
 -66 \hat R \hat R_{ij} + 108 \hat R^{l}{}_{i} \hat R_{lj}
-20 \hat R_{ij;k}{}^{;k} + 5 \hat R_{;ij} \Bigg \} \, .
\eqno(3.12b)
$$
In comoving synchronous gauge, conservation of particle
number implies that $\rho \gamma^{1/2}$ is independent of
time, and hence the number density $\rho$ is given by
$$
\rho(\chi, q) = { 4 \over 3 } \ k^{1/2}(q) \
\gamma^{-1/2}(\chi, q) \, . \eqno(3.13)
$$
More details are given in Salopek \& Stewart (1992), CPSS and
Parry et al (1994).

The argument for using the `square' trick is heuristic, and
it can placed on firmer theoretical grounds by showing that it leads
to an exact solution of Einstein's equations.
Our gradient expansion solution is essentially
a Taylor series that is accurate for small $\chi$.
{\it A priori} there is no reason that it should work well in the regime
where pancake structures are forming. However, after choosing an
appropriate seed,
the expressions for the `square' agree precisely with the
exact Szekeres (1975) solution which describes axisymmetric
pancake formation (see CPSS). Hence, we have essentially
{\it fitted} 
the late-time behavior to the
exact solution. This method is similar to the application
of WKB approximation in quantum mechanics. There, one matches
two WKB solutions across the classical turn-around point by
using the exact Airy solution for a linear potential.

\vs
\centerline{\bf 3.4 Modeling the Observable Universe}
\vs

Further simplifications arise when we wish to model
our observable Universe. Although gravitational
radiation arising from inflation (Salopek 1992) can
indeed contribute significantly to
the cosmic microwave background anisotropy
observed by the COBE satellite, its interaction with cold-dark-matter
is negligible during the matter-dominated era.
For the purposes of describing the formation of
galaxies, it is adequate to write the initial
seed metric in the isotropic form:
$$
k_{ij}(q) = \delta_{ij} \, {\rm exp}
\left ( { 10 \over 3 } f(q) \right ) \ . \eqno(3.14)
$$
The arbitrary function $f(q)$ is Bardeen's gauge-invariant variable
at early times, and
it can be interpreted as the negative of the initial
gravitational potential.
Whereas all of the expressions in section 3.3 were invariant
under the group of spatial coordinate transformations, eq.(3.14) is
invariant only under the subgroup of spatial rotations.
We have thus sacrificed some of the gauge-invariance
in favor of expressing the seed-metric in a very simple
form involving the single function $f(q)$.
(However, in the derivation of eqs.(3.11), (3.12), it is very
useful  to assume that all quantities are
fully gauge-invariant, otherwise the expressions become unwieldy.)
One may readily compute the Ricci tensor for the above
seed metric:
$$
\hat R_{ij}(q) = { 5 \over3}
\left ( - f_{,ij} - \delta_{ij} f^{,l}{}_{,l} \right )+
{ 25 \over 9} \left ( f_{,i} f_{,j} - f_{,l} f^{,l}
\delta_{ij} \right ) \ . \eqno(3.15)
$$

Our results may be simplified even further if we
make one additional approximation: at each order in the
gradient expansion, we will retain only the
lowest order terms in $f$ since $f \sim 10^{-5}$
is much smaller than unity.  For example,
we obtain
$$
\gamma^{(1)}_{ij}(\chi, q)= \chi^{4/3} \delta_{ij}, \quad {\rm (first)}
\eqno(3.16)
$$
at first order,
$$
\tilde \gamma^{(3)}_{ij}(\chi, q)= \chi^{4/3}
\[ \delta_{il} + { 3\over 2 }  \chi^{2/3} f_{,il} \]
\[ \delta^l_j + { 3\over 2 }   \chi^{2/3} f^{,l}{}_{,j} \],
\quad {\rm (improved \ 3rd)}
\eqno(3.17)
$$
at improved third order, and
$$
\tilde \gamma^{(5)}_{ij}(\chi,q) = \chi^{4/3}
\[ \delta_{il} + { 3 \over 2 } \chi^{2/3}
f_{,il} + \chi^{4/3} F_{il} \]
\[ \delta^l_j + { 3 \over 2 } \chi^{2/3}
f^{,l}{}_{,j} + \chi^{4/3} F^l{}_j \]
\quad {\rm (improved \ 5th)} \eqno(3.18a)
$$
at improved fifth order
where
$$
F_{ij} = { 27 \over 56} \left [
4 \left ( f_{,il} f^{,l}{}_{,j} - f^{,l}{}_{,l} f_{,ij} \right )
+ \delta_{ij}
\left ( (f^{,l}{}_{,l})^2 - f^{,lm} f_{,lm} \right ) \right ]
\ . \eqno(3.18b)
$$
It is important to note that the number density at the
improved third order is
exactly the result quoted by Zel'dovich:
$$
\tilde \rho^{(3)}(\chi, q) = { 4 \over 3 \chi^2} \ /
\text{det} \[ \delta^i{}_l + { 3 \over 2 } \chi^{2/3} f^{,i}{}_{,l}(q) \].
\eqno(3.19)
$$
Hence we are justified in claiming that our
improved third order spatial gradient approximation
reproduces  the usual Zel'dovich approximation
(CPSS, Salopek et al (1994)).

\vs
\centerline{\bf 3.5 Calculation of Collapse Epoch }
\vs
In order to understand the consequences of the improved
fifth order results, we will compute the time when
the density becomes infinite.

For a given spatial point, we will rotate
our coordinates so that the matrix of second derivatives
is diagonal,
$$
f_{,ij} \equiv {\partial^2 f \over \partial q^i \partial q^j }=
\text{diag} [ -d_1, -d_2, -d_3 ] \ ,  \eqno(3.20)
$$
Fortunately, the 3-metric at improved
fifth order, eq.(3.18), is also diagonal, which
gives it a huge advantage over the original form eq.(3.12).
The density $\tilde \rho^{(5)}$ at improved fifth order
becomes infinite when
$$
\text{det }\Bigg [ \delta_{ij} + { 3 \over 2 } \chi^{2/3} f _{,ij}
+ { 27 \over 56} \chi^{4/3} \left [
4 \left ( f_{,il} f^{,l}{}_{,j} - f^{,l}{}_{,l} f_{,ij} \right )
+ \delta_{ij}
\left ( (f^{,l}{}_{,l})^2 - f^{,lm} f_{,lm} \right ) \right ]
\Bigg ] =0. \eqno(3.21)
$$
If $d_1$ is the  largest positive eigenvalue of $-f_{,ij}$,
then this occurs when the scale factor ($a\equiv \chi^{2/3}$) is
given by
$$
a^{(5)}_{coll} = { 4 \over 3 d_1} /
\left [  1 + \sqrt{ 1 + { 12 \over 7 }
( { d_3 \over  d_1} + { d_2 \over d_1} -
{ d_2 \over d_1}  { d_3 \over  d_1}  )} \right ], \eqno(3.22)
$$
which is a function of $q$.
In Fig.~1 we plot the collapse time for various
values of the order pair $(d_2/d_1,  d_3/d_1)$
which describe various ellipsoid
configurations of the quadratic form
$f_{,ij} \Delta q^i \Delta q^j=1$.
(1,1) is a sphere, (1,0) and (0,1) are cylinders,
and the line $d_2/ d_1 = d_3/ d_1$ represents a class of
oblate spheroids.

The origin (0,0) corresponds to the situation
when $d_1 >> |d_2|, |d_3|$ which describes
a sheet or plane. As expected from the discussion
at the end of section 2.2,
the standard Zel'dovich approximation (improved 3rd order)
is essentially exact for this case.
However, when all the eigenvalues
are equal,  $d_1 =d_2 = d_3$,
describing a locally spherical system, then
$a^{(5)}_{coll}/ a^{(3)}_{coll} = 0.755$ represents a
$24\%$ decrease over the improved third order result.
The fifth order correction can be quite significant
although it probably breaks down before one reaches
$a^{(5)}_{coll}/ a^{(3)}_{coll} = 2$ shown in Fig.~1.

\vs
\centerline{\bf 4. \quad COMPARISON WITH EXACT SOLUTIONS OF EINSTEIN'S
EQUATIONS }
\vs
In CPSS, we compared our improved spatial gradient expansion
with the Szekeres solution. We will now compare with two
well-known exact solutions: planar and spherical geometries.

\vs
\centerline{\bf 4.1 \quad Planar Geometries and the
Zel'dovich Approximation}
\vs
For a planar gravitational system, the Zel'dovich approximation
is essentially exact until caustic formation
provided that the {\it initial}
gravitational potential $ f(z)$ satisfies,
$$
f(z) \ll 1,  \eqno(4.1a)
$$
and
$$
\left( { \partial  f \over \partial z} \right )^2  \ll
{ \partial^2  f \over \partial z^2} ,
\eqno(4.1b)
$$
where $z$ denotes the spatial coordinate.
This result is well-known for the Newtonian theory
(see, {\it i.e.}, Kofman (1991))
but here we will use the methods of general relativity.
For a perturbation $ f(z) =  f_0 \, {\rm cos}(kz)$,
with $f_0 < 10^{-5}$ consistent with observations,
condition (4.1a) implies condition (4.1b),
but to be precise and mathematically
rigorous, we will assume that both conditions hold.

The metric for a planar gravitational system was given in the
following form by Eardley, Liang \& Sachs (1972):
$$
ds^2 = -d\chi^2+ \phi^2(\chi, z) \left( dx^2 + dy^2 \right )
+ \psi^2(\chi, z)  dz^2 \eqno(4.2a)
$$
where
$$
\left ( { d \phi \over d \chi} \right )^2 =
{ \lambda(z) \over \phi} + \kappa^2(z) \, , \eqno(4.2b)
$$
and
$$
\psi= { \partial \phi \over \partial z} / \kappa(z) \, . \eqno(4.2c)
$$
Here $\lambda \equiv \lambda(z)$,  $\kappa \equiv \kappa(z)$
are independent of time.

If $\lambda > 0$ (which is the case of physical interest ---
see below), then the exact solution is given
parametrically in terms of $\eta$,
$$
\chi = { \lambda \over 2 \kappa^3} \left ( {\rm sinh} \eta - \eta \right )
\, , \eqno(4.3a)
$$
$$
\phi = { \lambda \over 2 \kappa^2} \left ( {\rm cosh} \eta - 1 \right )
\, , \eqno(4.3b)
$$
$$
\psi = {\lambda \over 2 \kappa^3} \left[
\left ( {\lambda^\prime \over \lambda} - 2 {\kappa^\prime \over \kappa} \right)
\left( {\rm cosh} \eta - 1 \right ) +
\left ( -{\lambda^\prime \over \lambda} +3 {\kappa^\prime \over \kappa} \right)
{\rm sinh} \eta  \,
{ \left( {\rm sinh} \eta - \eta  \right) \over
\left ( {\rm cosh} \eta - 1 \right )}  \right ]\, , \eqno(4.3c)
$$
where prime denotes a derivative with respect to $z$:
$\; {}^\prime= \partial / \partial z$.  It is useful to let
$$
\lambda = { 4 \over 9 } {\rm exp} \left ( 5 f \right ) \, , \quad
\kappa =  {5 f^\prime \over 3 },     \eqno(4.4)
$$
because at early times (small $\chi$), one recovers the
standard long-wavelength initial conditions,
$$
\phi = \psi =\chi^{2/3} \, {\rm exp} \left( { 5 f \over 3} \right ) \,
\quad ({\rm early \; times}) \, ,\eqno(4.5)
$$
which describes an initially isotropic metric.
The metric variables become
$$
\chi = { 6 \over 125 }
{ {\rm exp} ( 5 f) \over \left( f^\prime\right )^3 }
\left ( {\rm sinh} \eta - \eta \right ) \, , \eqno(4.6a)
$$
$$
\phi =
{ 2 \over 25 }
{ {\rm exp} ( 5 f) \over \left( f^\prime\right )^2 }
\left ( {\rm cosh} \eta - 1 \right ) \, , \eqno(4.6b)
$$
$$
\psi = { 6 \over 125 }
{ {\rm exp} ( 5 f) \over \left( f^\prime\right )^3 }
\left[
\left ( 5 f^\prime - 2 { f^{\prime \prime} \over
f^\prime   } \right)
\left ( {\rm cosh} \eta - 1 \right ) +
\left ( -5 f^\prime +3  { f^{\prime \prime} \over
 f^\prime  } \right)
{\rm sinh} \eta  \,
{ \left ( {\rm sinh} \eta - \eta  \right ) \over
\left ({\rm cosh} \eta - 1 \right )}
\right] \, . \eqno(4.6c)
$$

Following Tomita (1975), we expand our results in
a Taylor series in $\chi^{2/3}$:
$$
\eqalignno{
\eta = 5 \chi^{1/3} \, f^\prime \,
{\rm exp} \left( -{ 5 f\over 3 }  \right ) \,
\Biggl [ 1 - { 5 \over 12 } \chi^{2/3} & \left (f^\prime \right )^2
{\rm exp} \left( -{ 10 f  \over 3 } \right ) \,+ \cr
{ 25 \over 56}  \chi^{4/3} &\left (f^\prime \right )^4
{\rm exp} \left( -{20 f  \over 3 }  \right )  + \ldots \Biggr ], &(4.7)
\cr }
$$
$$
\phi = \chi^{2/3}
{\rm exp} \left( { 5 f \over 3 }  \right ) \,
\left [ 1 + { 5 \over 4 } \chi^{2/3} \left (f^\prime \right )^2
{\rm exp} \left( -{ 10 f \over 3 }  \right ) \,
- { 75 \over 112} \chi^{4/3} \left (f^\prime \right )^4
{\rm exp} \left( -{20 f \over 3 }  \right ) \ldots \right ], \eqno(4.8)
$$
$$
\eqalignno{
\psi &= \chi^{2/3}
{\rm exp} \left ( { 5 f \over 3 }  \right ) \,
\biggl[    1 + \left( { 3 \over 2 } f^{\prime \prime}
- { 5 \over 4}  \left (f^\prime \right )^2\right ) \chi^{2/3}
{\rm exp} \left( -{ 10 f \over 3 }  \right ) \,+ \cr
& \left( -{45 \over 28}  \left (f^\prime \right )^2  f^{\prime \prime}
+ {225 \over 112}  \left (f^\prime \right )^4\right )
\chi^{4/3}
{\rm exp} \left( -{20 f \over 3 }  \right ) + \ldots \biggr ]  .
&(4.9)\cr }
$$
It can be shown that eqs.(4.8) and (4.9) are in exact agreement
with the improved spatial gradient expansion given in eq.(3.12).

Let's now focus our attention on the last equation, and
apply the approximations of eq.(4.1) which leads to:
$$
\psi = \chi^{2/3}
\left [ 1 +  { 3 \over 2 } f^{\prime \prime} \chi^{2/3}
-{45 \over 28} \left (f^\prime \right )^2  f^{\prime \prime}
\chi^{4/3} + \ldots \right ] \, . \eqno(4.10)
$$
When caustic formation occurs, the second term is $-1$
and the third term is negligible provided eq.(4.1b) is valid.
(All higher order terms are negligible as well.)
To an excellent approximation, the fields evolve according to
$$
\eta = 5 f^\prime \chi^{1/3} \, ,\eqno(4.11)
$$
$$
\phi= \chi^{2/3} \, , \eqno(4.12)
$$
$$
\psi= \chi^{2/3} \left( 1 + { 3 \over 2 } \chi^{2/3} f^{\prime \prime}
 \right )\, , \eqno(4.13)
$$
before caustic formation.
The last expression is the Zel'dovich approximation because the
corresponding density is
$$
\rho = { 4 \over 3 \chi^2 }  { 1 \over
\left | 1 + { 3 \over 2 } \chi^{2/3} f^{\prime \prime}  \right | }\, .
\eqno(4.14)
$$

Hence in a planar system, the Zel'dovich approximation is
essentially exact until
caustics form. Additional boundary conditions are needed to
describe what happens later. Two possibilities are:

(1) The particles are collisionless and they pass through
each other. (These are cold-dark-matter particles;
they comprise approximately 95\% of the mass density of the Universe.)

(2) The particles stick together on collision, forming
a single particle whose velocity is given by momentum
conservation. This presumably describes the baryon component (which
comprises about 5\% of the mass density of the Universe.)

We will not discuss the issue of caustics further in this paper.

\vs
\centerline{\bf 4.2 \quad Spherical Systems and the
Improved Zel'dovich Approximation}
\vs
For spherical systems, the usual Zel'dovich approximation
may not be very accurate. As a result, we investigate higher
order corrections.

The Tolman-Bondi metric is
$$
ds^2 = -d\chi^2+
r^2(\chi, q) \left( d\theta^2 + {\rm sin}^2 \theta d \phi^2 \right )
+ \psi^2(\chi, q)  dq^2, \eqno (4.15)
$$
where
$$
\left( {d r \over d \chi}\right )^ 2 =
{2M  \over r} + \left[ \Gamma^2(q) -1 \right] , \eqno(4.16a)
$$
$$
\psi= { \partial r \over \partial q} / \Gamma(q),\eqno(4.16b)
$$
and $M \equiv M(q)$,  $\Gamma \equiv \Gamma(q)$
are functions of the comoving radial coordinate $q$;
they are independent of time $\chi$. (See Misner et al (1973).)

We now choose
$$
M = { 2 \over 9 } q^3 {\rm exp}\left( 5 f(q) \right ),
$$
$$
\Gamma= 1 + { 5\over 3 } q {\partial f \over \partial q}  ,
$$
so that for small $\chi$, the
metric variables evolve according to
$$
r/ q = \psi =\chi^{2/3} \, {\rm exp} \left( { 5 f \over 3} \right ) \,
 , \eqno(4.17)
$$
which yields an isotropic metric at early times.

For $\Gamma^2 -1 > 0$,
the exact solution is analogous to the planar case
$$
\chi = { 2 \over 9} { {\rm exp} \left(5 f \right) q^3 \over
\left ( \Gamma^2 -1 \right )^{3/2} }
\left ( {\rm sinh} \eta - \eta \right ) \, , \eqno(4.18a)
$$
$$
r = { 2 \over 9} { {\rm exp} \left(5 f \right) q^3 \over
\left ( \Gamma^2 -1 \right ) }
\left ( {\rm cosh} \eta - 1 \right ) \, . \eqno(4.18b)
$$
If $\Gamma^2 -1 < 0 $, then the above formulae are still
valid; in this case $\sqrt{\Gamma^2 -1}$ and
$\eta$ are pure imaginary.

The planar case may be recovered if one identifies
$\phi= r/ q$, and then assumes
$$
q { \partial f \over \partial q} \gg 1;  \eqno(4.19)
$$
crudely speaking, this is the large radius limit.
However, we have chosen to treat the spherically
symmetric case separately because the results
are qualitatively different.

Once again, we expand in a Taylor series in $\chi^{2/3}$
$$
\eqalignno{
r& = \chi^{2/3}
q \, {\rm exp} \left( { 5 f \over 3 } \right )  \,
\Biggl \{ 1 +  { 5 \over 4 } \chi^{2/3}
\left[ \left(  { \partial f \over \partial q} \right )^2 +
{ 6 \over 5 q} { \partial f \over \partial q} \right ]
{\rm exp} \left( -{ 10 f \over 3 }  \right ) \, \cr
&- { 75 \over 112} \chi^{4/3}
\left[ \left( { \partial f \over \partial q} \right )^2 +
{ 6 \over 5 q} { \partial f \over \partial q} \right ]^2
{\rm exp} \left( -{20 f \over 3 }  \right )  \cr
&+ { 2875 \over 4032}  \chi^2
\left[ \left( { \partial f \over \partial  q} \right )^2 +
{ 6 \over 5 q} { \partial f \over \partial q} \right ]^3
{\rm exp} \left( -{10 f }  \right )
+ \ldots \Biggr \} \, ,  &(4.20a) \cr}
$$
$$
\eqalignno{
\psi& =   \chi^{2/3}
{\rm exp} \left ( { 5 f \over 3 }  \right ) \,
\Biggl \{  1 +  \chi^{2/3} \left[ { 3 \over 2 }
{\partial ^2 f \over \partial q^2}
- { 5 \over 4}  \left ( { \partial f \over \partial q}
\right )^2 \right ]
{\rm exp} \left( -{ 10 f \over 3 }  \right ) \,  &(4.20b) \cr
& - {45 \over 28} \chi^{4/3}
\left[ \left( { \partial f \over \partial q} \right )^2 +
{ 6 \over 5 q} { \partial f \over \partial q} \right ]
\left[ { \partial ^2 f \over \partial q^2} -
{ 1 \over 2 q} {\partial f \over \partial q}
-{ 5 \over 4}  \left( {\partial f \over \partial q} \right )^2 \right ]
{\rm exp} \left( -{20 f \over 3 }  \right )     \cr
&+ { 575 \over 224}  \chi^2
\left[ \left( { \partial f \over \partial q} \right )^2 +
{ 6 \over 5 q} { \partial f \over \partial q} \right ]^2
\left[ { \partial^2 f \over \partial q^2} - {2 \over 3 q}
{\partial f \over \partial q}
-{ 25 \over 18}  \left( {\partial f \over \partial q} \right )^2 \right ]
{\rm exp} \left( -10 f   \right )  + \ldots \Biggr \}
\, .   \cr }
$$
(Of course, eqs.(4.20) are in exact agreement with the
improved spatial gradient expansion eq.(3.12).)
In general, this series does not truncate after the first
two terms. For example, consider $ f(q) =  f_0 \cos(Aq) $
at $q=0$.

In Fig.~2a, for $\Gamma^2 -1 < 0$,
we show the exact solution for the radius $r(\chi, q)$  (bold line) with $q=0$.
This case describes the collapse of a high density region
which is spherically symmetric.
The light line is the first term (long-wavelength approximation).
Also shown are the terms of order 3, 5 and 7 in spatial gradients.
It is encouraging
that the series appears to be converging before total
collapse occurs. In Fig.~2b
we consider $\Gamma^2 -1 > 0$ describing the evolution of
a void region.  Here, the series does not
converge for $ \chi > 1.4$ (which is also the radius
of convergence of Fig.~2a). Even then the higher order
terms are relatively close to the exact result.


\vs
\centerline{\bf 5.  \quad COMPARISON WITH THE BUCHERT-EHLERS SCHEME}
\vs

Various schemes offering higher order improvements to the Zel'dovich
approximation within Newtonian theory have been suggested, including
Buchert (1989, 1992, 1993), Buchert \& Ehlers (1993), Moutarde et
al (1991), Gramann (1993), among others.
Naturally one would like to be able to compare such theories with the
one presented here.
We have concentrated on the Buchert-Ehlers second order scheme
(which we derived in section 2),
because their theory is clearly and unambiguously expressed.
Although there is a related third order scheme (Buchert 1993), the
calculations required for a comparison are too unwieldy to discuss
here.

All Newtonian theories are non-local because at some stage the
Newton-Poisson equation has to be solved.
We therefore  discuss only the restricted model described in section 5
of their paper which depends on the scalar function $f(q)$
which obeys the constraint
$$
\epsilon _{ijk} f_{,j}f_{,kmm} = 0 \, . \eqno(5.1)
$$
For such a choice of $f(q)$, the non-local Newtonian
expression (2.17) becomes local:
$$
x_{(B)}^i = q^i + \threehalves t^{2/3}f_{,i} +
\textstyle\frac{27}{56}t^{4/3}
\(f_{,ik}f_{,k} - f_{,kk} f_{,i}\). \eqno(5.2)
$$
Working in synchronous gauge their approximation implies a spatial
$3$-metric
$$
\eqalignno {
\gamma _{(B)ij}
=& t^{4/3}\,\Partial{x_{(B)l}}{q^i}\;\Partial{x_{(B)}^l}{q^j} \cr
=& t^{4/3}\(\delta _{il}+\threehalves t^{2/3} f_{,il} +
t^{4/3}{\cal F}_{il}\)
\(\delta ^l{}_j + \threehalves t^{2/3} f^{,l}{}_{,j} +
t^{4/3}{\cal F}^l{}_j\), & (5.3)
}
$$
where
$$
{\cal F}_{ij} =
\textstyle\frac{27}{56}
\(f_{,ijk} f^{,k} + f_{,ik} f_{,j}{}^k -
f^{,k}{}_{,(i} f_{,j)k} -
f^{,k}{}_{,k} f_{,ij}\). \eqno(5.4)
$$
If we set $t = \chi$ and ignore ${\cal F}_{ij}$,
then their resulting first order theory
agrees perfectly with our improved third order theory, eq.(3.17).
Both theories imply the standard Zel'dovich approximation
expression for the density, eq.(3.19).
However their second order theory (${\cal F}_{ij}\ne 0$) does not
in general agree with our improved fifth order approximation, eq.(3.18).
Their ${\cal F}_{ij}$  is not the same as our $F_{ij}$, eq.(3.18b).
In particular the geodesics, which relate Eulerian to Lagrangian
positions as a function of time, will in general differ.

For spherical geometries, the two expressions,
${\cal F}_{ij}$  and $F_{ij}$, are indeed
consistent. Moreover, it is straightforward to verify
in general, that
their traces coincide: ${\cal F}^k{}_k = F^k{}_k$.
This implies that
$\text{det}\gamma _{(B)ij} = \text{det}\tilde\gamma ^{(5)}_{ij}$:
at this order, both estimates of the density as a
function of $t$ and $q$ will agree.
The discrepancy between  ${\cal F}_{ij}$  and $F_{ij}$
is still under investigation.


\vs
\centerline{ \bf 6. \quad DISCUSSION AND CONCLUSIONS }
\vs

Since the velocity potential plays such a prominent
role in the Zel'dovich approximation, we began our
analysis with a relativistic action principle for
a dust field $\chi$. In the Newtonian limit, we used
the {\it non-relativistic HJ equation} to solve for the
velocity potential which is expressed in terms of a
Taylor series. We recovered the results of
Buchert \& Ehlers (1993). Unfortunately the final expressions for
the density are nonlocal, which make further analytic progress
difficult.

In conflict with the principle of causality,
nonlocal terms suggest that information may be transferred faster
than the speed of light. By considering a general
relativistic formulation, we avoid this problem.
We solved Einstein's equations
using a spatial gradient approximation. We
derived the Zel'dovich approximation
using the first two terms in our expansion.
We also investigated higher order corrections.
Consistent with causality,
our expressions are local in that they involve only derivatives
of the initial gravitational potential. The simplicity of our final
results suggest that they will be a useful tool in cosmology.
For arbitrary initial conditions, we  computed the
epoch of collapse. We hope to apply our techniques
to the interpretation of redshift surveys.

As a check, we have compared our gradient expansion
with two exact solutions of Einstein's equations: planar
and spherical geometries. Using a Taylor series in
the planar case, it
is straightforward to show that the first two terms
in the improved expansion
are sufficient --- all the higher order terms are very
small. Hence in a general relativistic formulation,
we rederive the fact that the Zel'dovich approximation
is essentially exact for planar geometries
(for a Newtonian derivation, see, e.g., Kofman (1991)).
As was expected from the Newtonian theory, the results for
spherical geometries differ quantitatively from the planar case.
Near zero radius, the series does not truncate for
a spherical system, and the higher corrections to
the Zel'dovich approximation are indeed significant.
(However, in the large radius limit the spherical
case reduces to the planar one.)
Our general relativistic approach, of, course is valid in general ---
not just for spherical and planar geometries.

Comparison with Newtonian theories, e.g., that of Buchert and
Ehlers (1993), is difficult because they are non-local.
For the special case of spherical geometries, our
results are completely consistent. However,
in general  we have not been able to show that two methods
are equivalent. In any case, our relativistic formalism does lead to
simpler expressions for the final results (see eq.(3.18)).

\vs
\centerline{\bf ACKNOWLEDMENTS}
\vs
D.S.S. received partial support from the Natural Sciences and
Engineering Research Council of Canada, and the
Canadian Institute for Theoretical Astrophysics
(CITA National Fellowship). K.M.C. was supported by the
Science and Engineering Research Council of the UK.

\vs
\centerline{\bf REFERENCES}
\vs

\noindent
Bardeen, J.M., 1980, Phys. Rev. D,  22, 1882

\noindent
Bertschinger, E. \& Jain, B., 1993, MIT preprint

\noindent
Bouchet, F.R., Juszkiewicz, R., Colombi, S. \& Pellat, R.,
1992, ApJ, 394, L5

\noindent
Broadhurst, T.J., Ellis, R.S.,  Koo,  D.C. \&  Szalay, A.S., 1990,
Nature, 343, 726

\noindent
Buchert, T., 1989, A\&A, 223, 9

\noindent
Buchert, T., 1992, MNRAS, 254, 729

\noindent
Buchert, T., 1993, MNRAS, (in press)

\noindent
Buchert, T \& Ehlers, J., 1993, MNRAS, 264, 375

\noindent
Comer, G.L., Deruelle, N., Langlois, D. \& Parry, J., 1994, Meudon preprint

\noindent
Croudace, K.M.,  Parry, J.,  Salopek, D.S. \& Stewart, J.M.,
1994, ApJ,  (in press)

\noindent
Dekel, A., Bertschinger, E. \& Faber, S.M., 1990,
ApJ, 364, 349

\noindent
De Lapparent, V., Geller, M.J., \& Huchra, J.P., 1986,
ApJ, 302, L1

\noindent
Ehlers, J., 1961, Abh. Math-Naturwiss. Klasse der Akad. Wiss.
und Lit., Mainz, nr. 11

\noindent
Eardley, D., Liang, E. \& Sachs, R., 1972, J. Math. Phys.,
13, 99

\noindent
Goldstein, H., 1981, Classical Mechanics,  Addison Wesley

\noindent
Gramann, M., 1993, ApJ, 405, L47 

\noindent
Kofman, L., 1991, in Sato, K.,  ed., Proc. IUPAP Conf.
Primordial Nucleosynthesis and Evolution of the
Early Universe, p. 495,  Kluwer, Dordrecht

\noindent
Landau, L. \& Lifshitz E.M., 1975, The Classical Theory of
Fields, (fourth English edition) Pergamon (Oxford)

\noindent
Lifshitz, E.M. \& Khalatnikov, I.M., 1964,
Usp. Fiz. Nauk, 80,
391 [Sov. Phys. Usp.,  6, 495 (1964)]

\noindent
Matarrese, S., Pantano, O. \& Saez, D., 1993, Phys. Rev. D,
47, 1311

\noindent
Misner, C.W., Thorne, K.S. \& Wheeler, J.A., 1973, Gravitation,
Freeman (New York)

\noindent
Moutarde, F., Alimi, J.-M., Bouchet, F.R., Pellat, R. \&
Ramani, A., 1991, ApJ, 382, 377

\noindent
Nusser, A., Dekel, A., Bertschinger, E. \&  Blumenthal, G.R., 1991,
ApJ, 379, 6

\noindent
Parry, J., Salopek, D.S. \& Stewart, J.M., 1994,
Phys. Rev. D (in press)

\noindent
Peebles, P.J.E., 1980, Large Scale Structure of the
Universe, (Princeton University Press)

\noindent
Peres, A.,  1962, Nuovo Cim., 26, 53

\noindent
Salopek, D.S., 1992, Phys. Rev. Lett.,  69, 3602

\noindent
Salopek, D.S. \& Stewart, J.M., 1992,
Class. Quantum Grav.,  9, 1943

\noindent
Salopek, D.S. \& Stewart, J.M., 1993, Phys. Rev. D,
47, 3235

\noindent
Salopek, D.S., Stewart, J.M., Croudace, K.M. \& Parry, J., 1994,
in Sato, K., ed., Proc. of 37th Yamada Conference,
Evolution of the Universe and its Observational Quest,
Tokyo, Japan, (Universal Academic Press)

\noindent
Schutz, B.F., 1970, Phys. Rev. D, 2, 2762

\noindent
Schutz, B.F., 1971, Phys. Rev. D,  4, 3559

\noindent
Shandarin, S.F. \& Zel'dovich, 1989, Rev. Mod. Phys.,
61, 185

\noindent
Stewart, J.M., 1990, Class. Quan. Grav., 7, 1169

\noindent
Szekeres, P., 1975, Commun. Math. Phys., 41, 55

\noindent
Tomita, K., 1975, Prog. Theor. Phys., 54, 730

\noindent
Zel'dovich, Ya.B., 1970, A\&A 5, 84
\vfill\eject

\vs
\centerline{\bf Figure Captions}
\vs
\noindent
{\bf Figure 1.} Higher order corrections to the
standard Zel'dovich approximation are in general significant.
For the improved fifth order calculation, the scale factor,
$a^{(5)}_{coll}\equiv \chi^{2/3}$, at the time of
collapse is shown as a function of the
eigenvalues $d_i$ of $-f_{,ij}$ where
$-f$ is the initial gravitational potential.
 From the bottom to the top, the curves  represent
$a^{(5)}_{coll}/ a^{(3)}_{coll} = 2.0, 1.2, 1.0 \ {\rm (bold)}, 0.9, 0.8
\ {\rm and} \ 0.755$ (sharp corner), where the first and last  are the
maximum and minimum values, respectively.
$d_1$ is the largest positive eigenvalue and
$a^{(3)}_{coll} = 2/(3 d_1)$ is the standard
Zel'dovich approximation result.

\vs
\noindent
{\bf Figure 2.} We compare our gradient expansion
(1st, 3rd, 5th and 7th order) with the
exact Tolman-Bondi solution (bold curve) for a spherical system.
In (a), spherical collapse is shown. Apparently,
the spatial gradient expansion is converging. In (b), the radius
$r$ expands indefinitely. The radius of convergence
$\chi_{crit}= 1.4$ is finite and it
coincides with the collapse time given in (a).
Even beyond the radius of convergence, the
successive approximations do not differ much from the
exact solution.

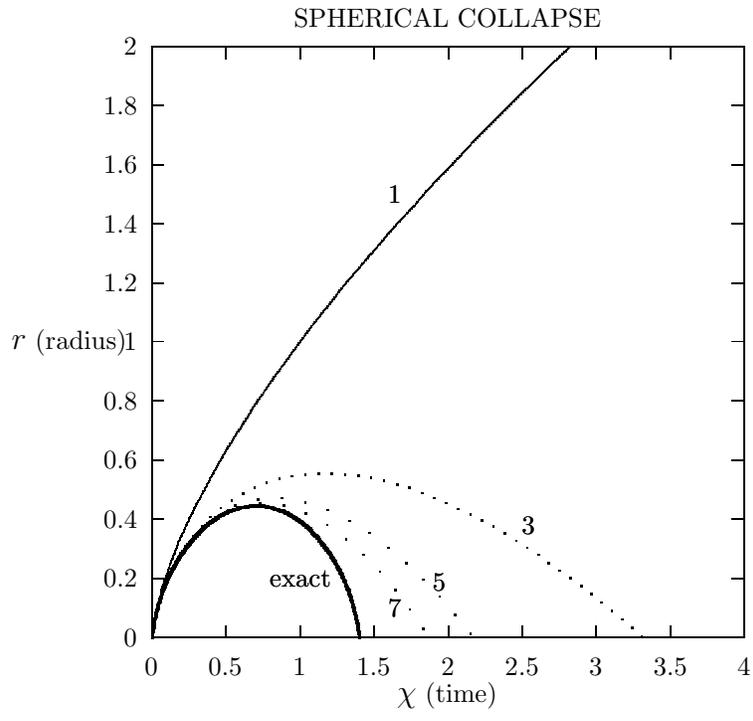
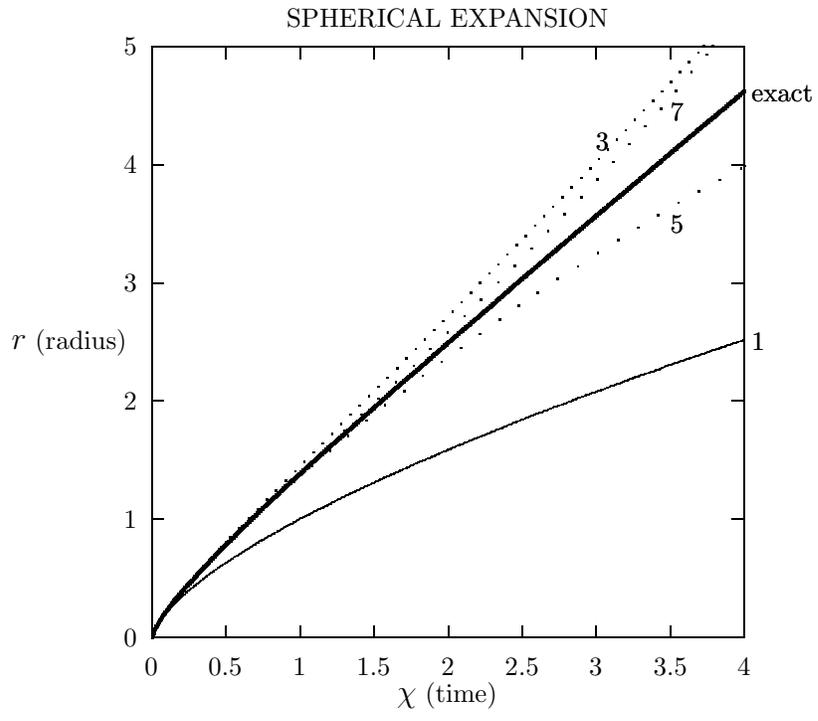
\bye